\documentclass[10pt, letterpaper, twocolumn]{article}
\usepackage{usenix}
\usepackage{citesort}
\usepackage{amssymb,graphicx,floatflt,url,bcprules,listings,
        textcomp,stmaryrd,xypic,calc,cancel,amsmath,mathpartir,tweaklist,skull}

\lstdefinelanguage{LambdaJS}{
sensitive=false,
morekeywords={let,func,return,undefined,null,typeof,true,false,ref,deref,
              var,new,break,while,try,catch,finally,throw,err,if,else,for,do,
              with,delete,new,instanceof,tagcheck,setref,in,tagerr,cheat},
keywordstyle=\bfseries\sffamily,
identifierstyle=\sffamily,
comment=[l]{//},
commentstyle=\sffamily,
string=[d]{"},
stringstyle=\ttfamily,
mathescape=true,
extendedchars=true,
basicstyle=\footnotesize\ttfamily,
showstringspaces=false,
numbers=none,
firstnumber=1,
numberstyle=\tiny,
stepnumber=1,
numbersep=5pt,
upquote=true,
columns=fullflexible,
flexiblecolumns=true,
}

\lstdefinelanguage{JavaScript}{
sensitive=false,
morekeywords={let,function,return,undefined,null,typeof,true,false,ref,deref,
              var,new,break,while,try,catch,finally,throw,err,if,then,else,for,do,
              with,delete,new,instanceof,this,},
keywordstyle=\bfseries\ttfamily,
identifierstyle=\ttfamily,
comment=[l]{//},
commentstyle=\ttfamily,
string=[d]{"},
stringstyle=\ttfamily,
mathescape=true,
extendedchars=true,
basicstyle=\footnotesize\ttfamily,
showstringspaces=false,
numbers=none,
firstnumber=0,
numberstyle=\tiny,
stepnumber=5,
numbersep=5pt,
upquote=true,
columns=fixed,
flexiblecolumns=true,
}
\newcommand{\js}[2][LambdaJS]{\textrm{\lstinline[language=#1]|#2|}}
\newcommand{\code}[1]{\js[JavaScript]{#1}}

\newcommand{\ljs}{$\lambda_{JS}$}

\newcommand{\jscript}{JavaScript}
\newcommand{\as}{ADsafe}
\newcommand{\asjs}{{\tt\small adsafe.js}}
\newcommand{\lint}{JSLint}
\newcommand{\asty}{ADsafety}
\newcommand{\fbjs}{{\sc fbjs}}

\newcommand{\css}{{\sc css}}
\newcommand{\dom}{{\sc dom}}
\newcommand{\html}{{\sc html}}

\def\qed{\hfill$\blacksquare$}

\newcommand{\type}[1]{\textrm{{\small\sffamily #1}}}

\def\tlint{\type{Widget}}
\def\twindow{\type{Global}}
\def\tdom{\type{HTML}}
\def\badtype{\skull}
\newcommand{\tarray}[1]{\type{Array}$\langle$#1$\rangle$}

\def\tabsent{\type{Absent}}

\def\codep{\textit{code}}
\def\protop{\textit{proto}}

\def\fields{\textit{fields}}

\newcommand{\judge}[1]{\mbox{\sc #1}}

\def\authors{
  {\rm Joe Gibbs Politz}
  \quad
  {\rm Spiridon Aristides Eliopoulos}
  \quad
  {\rm Arjun Guha}
  \quad
  {\rm Shriram Krishnamurthi}
  \\\\
  Brown University
}

\newcommand{\figref}[1]{figure~\ref{#1}}
\newcommand{\Figref}[1]{Figure~\ref{#1}}

\newcommand{\secref}[1]{section~\ref{#1}}
\newcommand{\Secref}[1]{Section~\ref{#1}}

\newcommand{\defref}[1]{definition~\ref{#1}}

\newcommand{\Claimref}[1]{Claim~\ref{#1}}

\renewcommand{\enumhook}{\setlength{\topsep}{1pt}
  \setlength{\itemsep}{-2pt}}
\renewcommand{\itemhook}{\setlength{\topsep}{2pt}
  \setlength{\itemsep}{-2pt}}

\newtheorem{adsafety}{Definition}
\newtheorem{desugarclaim1}{Claim}
\newtheorem{desugarclaim2}[desugarclaim1]{Claim}
\newtheorem{jslint}[desugarclaim1]{Claim}

\newtheorem{linterface}{Lemma}
\newtheorem{preservation}[linterface]{Lemma}
\newtheorem{adsafe_no_dom}[linterface]{Lemma}
\newtheorem{adsafe_no_eval}[linterface]{Lemma}

\begin{document}

\title{ADsafety \\ \large{Type-Based Verification of JavaScript Sandboxing}}

\author{
   \authors
  }

\maketitle

\thispagestyle{empty} 

\begin{abstract}
Web sites routinely incorporate JavaScript programs from several sources into a
single page.  These sources must be protected from one another, which
requires robust sandboxing.  The many entry-points of
sandboxes and the subtleties of JavaScript demand robust verification
of the actual sandbox source.  We use a novel type system for
JavaScript to encode and verify sandboxing properties.  The resulting
verifier is lightweight and efficient, and operates on actual source.
We demonstrate the effectiveness of our technique by applying it to
ADsafe, which revealed several bugs and other weaknesses.
\end{abstract}

\section{Introduction}

A \emph{mashup} Web page displays content and executes
JavaScript from various untrusted sources.  Facebook applications,
gadgets on the iGoogle homepage, and various embedded maps are the
most prominent examples.  By now, mashups have become ubiquitous.
Indeed, web pages that display advertisements from ad networks are
also mashups, because they often employ JavaScript for animations and
interactivity.  A survey of popular pages shows that a large
percentage of them include scripts from a diverse array of external
sources~\cite{cy;hw:InsecureJSPractices}.  Unfortunately, these
third-party scripts run with the same privileges as trusted,
first-party code served directly from the originating site.  Hence,
the trusted site is susceptible to attacks by maliciously crafted
third-party software.

This paper addresses
language-based Web sandboxing systems,  one of several mechanisms for securing mashups.  Most sandboxing mechanisms have similar
high-level goals and designs, which we outline in \secref{sec:sandboxes}.  In
\secref{sec:codereview}, we review the design and implementation of
sandboxes and demonstrate the need for tool-supported verification.
\Secref{sec:plan} provides a detailed plan for the rest of the paper.
Our work makes several contributions:

\begin{enumerate}

\item A type system for general JavaScript programs, with support for
patterns found in sandboxing libraries;\footnote{See
  \url{cs.brown.edu/research/plt/dl/adsafety/v1} for our implementation, proofs,
and other details.}

\item a formal definition of safety properties for Yahoo!'s ADsafe
  sandbox in terms of this type system; and,

\item a type-based verification of the ADsafe framework, and
  descriptions of bugs and their fixes found while performing the
  verification.


\end{enumerate}

\section{Language-based Web Sandboxing\label{sec:sandboxes}}
\lstset{language=JavaScript}

\begin{figure}
  \centering
  \includegraphics{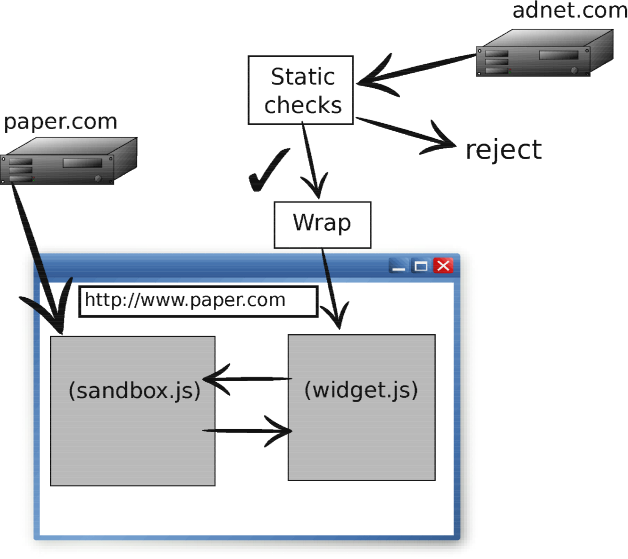}
  \caption{Web sandboxing architecture}
  \label{fig:sandbox-pic}
\end{figure}

The Web browser environment provides references to objects that
implement network access, disk storage, geolocation, and other
capabilities. Legitimate web applications use them for various reasons, 
but embedded widgets can exploit them because all
JavaScript on a page runs in the same global environment.  A Web
sandbox thus attenuates or prevents access to these capabilities,
allowing pages to safely embed untrusted widgets.
ADsafe~\cite{adsafe}, Caja~\cite{caja}, \fbjs~\cite{fbjs}, and
BrowserShield~\cite{cr;jd;hjw;od;se:BrowserShield} are
\emph{language-based} sandboxes that employ broadly similar security
mechanisms, as defined by Maffeis, et
al.~\cite{sm;jcm;at;FiltersRewritingWrappers}:
\begin{itemize}

\item  A Web sandbox includes a static code checker that
\emph{filters} out certain widgets that are almost certainly unsafe.
This checker
is run before the widget is delivered to the browser.

\item A Web sandbox provides runtime \emph{wrappers} that
attenuate access to the \dom{} and other capabilities.  These wrappers are
defined in a trusted runtime library that is linked with the untrusted
widget. 

\item Static checks are necessarily conservative and can reject benign
programs.  Web sandboxes thus specify how potentially-unsafe programs
are \emph{rewritten} to use dynamic safety checks.

\end{itemize}
This architecture is illustrated in \figref{fig:sandbox-pic}, where
 an untrusted widget from \url{adnet.com} is embedded in a page from
\url{paper.com}.  The untrusted widget is filtered by the static
checker.  If static checking passes, the widget is rewritten to invoke
the runtime library.  Both the runtime library and the checked, rewritten widget must be hosted on a site trusted by
\url{paper.com}, and are assumed to be free of tampering.

\paragraph{Reference Monitors} 
 A Web sandbox implements a
\emph{reference monitor}  between the
untrusted widget and the browser's capabilities.  Anderson's seminal
work on reference monitors identifies their certification
demands~\cite[p 10-11]{ja:ReferenceMonitor}:
\begin{quote}
The proof of [a reference monitor's] model security requires a
verification that the modeled reference validation mechanism is tamper
resistant, is always invoked, and cannot be circumvented.
\end{quote}
Therefore, a Web sandbox must come with a precisely stated notion
of security, and a proof that its static checks and runtime library
correctly maintain security.  The end result should be a quantified
claim of safety over \emph{all} possible widgets that execute against
the runtime library.

\section{Code-Reviewing Web Sandboxes\label{sec:codereview}}

Imagine we are confronted with a Web sandbox and asked to ascertain
its quality.  One technique we might employ is a code-review.
Therefore, we perform an imaginary review of a Web sandbox,
focusing on the details of ADsafe.  Later, we will discuss how to
(mostly) remove people from the loop.

ADsafe, like all Web sandboxes, consists of two interdependent
components:
\begin{itemize}

\item a static verifier, called JSLint,\footnote{JSLint can perform
other checks that are not related to ADsafe.  In this paper,
``JSLint'' refers to JSLint with ADsafe checks enabled.} which
filters out widgets not in a safe subset of JavaScript, and

\item a runtime library, \texttt{adsafe.js}, which implements
\dom{} wrappers and other runtime checks.

\end{itemize}
These conspire to make it safe to embed untrusted widgets, though
``safe'' is not precisely defined.  We will return to the definition
of safety in \secref{sec:plan}.

\paragraph{Attenuated Capabilities} 
Widgets should not be able to directly
reference various capabilities in the browser environment. Direct \dom{} references are particularly dangerous
because, from an arbitrary \dom{} reference, \lstinline|elt|, a widget can
simply traverse the object graph and obtain references to all
capabilities:
\begin{lstlisting}
var myWindow = elt.ownerDocument.defaultView;
myWindow.XMLHttpRequest;
myWindow.localStorage;
myWindow.geolocation;
\end{lstlisting}
Widgets therefore manipulate \emph{wrapped} \dom{} elements instead of
direct references.  \dom{} wrappers form the bulk of the runtime library
and include many dynamic checks and patterns that need to be verified:
\begin{itemize}

\item The runtime manipulates \dom{} references, but returns them to the widget
in wrappers.  We must verify that all returned values are in fact
wrapped, and that the runtime cannot be tricked into returning a direct \dom{}
reference.

\item The runtime calls \dom{} methods on behalf of the widget.  Many
methods, such as \lstinline|appendChild| and
\lstinline|removeChild|, require direct \dom{} references as
arguments. We must verify that the runtime cannot be tricked with
a maliciously crafted object that mimics the \dom{} interface and steals
references.

\item The runtime attaches \dom{} callbacks on behalf of the widget.
These callbacks are invoked by the browser with event arguments that
include direct \dom{} references.  We must verify that the runtime
appropriately wraps calls to untrusted callbacks in the widget.

\item The widget has access to a \dom{} subtree that it is allowed to manipulate.  
The runtime ensures that the widget only manipulates
elements in this subtree.  We must verify that various \dom{} traversal
methods, such
as \lstinline|document.getElementById| and \lstinline|Element.getParent|,
do not allow the widget obtain wrappers to elements outside its subtree.

\item The runtime wraps many \dom{} functions that are only conditionally safe.
For example, \lstinline|document.createElement| is usually safe,
unless it is used to create a \lstinline|<script>| tag, which can load
arbitrary code.  Similarly, the runtime may allow widgets to set \css{}
styles, but a \css{} {\sc url}-value can also load external code.  We must
verify that the arguments supplied to these \dom{} functions are safe.

\end{itemize}
ADsafe's \dom{} wrappers are called \emph{Bunches}, which wrap collections of \html{} elements.  
There are twenty Bunch-manipulating
functions that are exposed to the widget---in addition to several
private helper functions---that face all the issues enumerated above
and need to be verified.  These functions cannot be verified in
isolation, because their correctness is dependent on assumptions about
the kinds of values they receive from widgets.  These assumptions are
discharged by the static checks in JSLint and other runtime checks to
avoid loopholes and complexities in JavaScript's semantics.

\paragraph{JavaScript Semantics} 
A Web sandbox must contend with JavaScript features that hinder security:
\begin{itemize}

\item Certain JavaScript features are unsafe to use in widgets.  
For example, a widget can use \lstinline|this| to obtain
\lstinline|window|, so it is rejected by JSLint:
\begin{lstlisting}
f = function() { return this; };
var myWindow = f();
\end{lstlisting}
We must verify that the subset of JavaScript admitted by the static checker
does not violate the assumptions of the runtime library.

\item Many JavaScript operators and functions include implicit type
conversions and method calls that are difficult to reason about.  For
example, when an operator expects a string but is instead given an
object, it does not signal an error.  Instead, it calls the object's
\lstinline|toString| method.  It is easy to write a stateful
\lstinline|toString| method that returns different strings on different
calls.  Such an object can then circumvent dynamic safety checks that are
not carefully written to avoid triggering implicit method calls.
These implicit calls are avoided by carefully testing the runtime types
of untrusted values, using the \lstinline|typeof| operator.  Such tests
are pervasive in ADsafe.  As a further precaution, ADsafe tries to
ensure that widgets cannot define \lstinline|toString| and
\lstinline|valueOf| fields in objects.

\end{itemize}


\paragraph{JavaScript Encapsulation}

\begin{figure}
\begin{displaymath}
  \begin{array}{lcr}
    \textrm{ADSAFE} & : & \js{ADSAFE.get(obj,name)} \\
    \textrm{dojox.secure} & : & \js{get(obj,name)} \\
    \textrm{Caja} & : & \js{$\$$v.r($\$$v.ro('obj'), $\$$v.ro('name'))} \\
    \textrm{WebSandbox} & : & \js{c(d.obj,d.name)} \\
    \textrm{FBJS} & : & \js{a12345_obj[$\$$FBJS.idx(name)]} \\
  \end{array}
\end{displaymath}
\caption{Similar Rewritings for \js{obj[name]}}
\label{fig:checks}
\medskip\hrule
\end{figure}

JavaScript objects have no notion of private fields.  If object
operations are not restricted, a widget could access built-in
prototypes (via the \lstinline|__proto__| field) and modify the
behavior of the container.  Web sandboxes statically reject such
expressions:
\begin{lstlisting}
obj.__proto__;
\end{lstlisting}
There are various other dangerous fields that are also
\emph{blacklisted} and hence rejected by sandboxes.  However,
syntactic checks alone cannot determine whether computed field names
are unsafe:
\begin{lstlisting}
obj["__pro" + "to__"];
\end{lstlisting}
Widgets are instead rewritten to use runtime checks that restrict
access to these fields.  \Figref{fig:checks} shows the rewrites
employed by various sandboxes.  Some sandboxes insert these and other
checks automatically, giving the illusion of programming in ordinary
JavaScript; ADsafe is more spartan, requiring widget authors to insert
the dynamic checks themselves; but the principle remains the same.

Web sandboxes also simulate private fields with this method
by introducing fields and then preventing widgets from accessing them. 
For example,
ADsafe stores direct \dom{} references in the \lstinline|__nodes__| field
of Bunches, and blacklists the \lstinline|__nodes__| field.

\subsection*{The Reviewability of Web Sandboxes\label{sec:reviewability}}

We have highlighted a plethora of issues that a Web sandbox must
address, with examples from ADsafe.  Although ADsafe's source follows
JavaScript ``best practices,'' the sheer number of checks and
abstractions make it difficult to review.  There are approximately
$50$ calls to three kinds of runtime assertions, $40$ type-tests, $5$
regular-expression based checks, and $60$ \dom{} method calls in the
$1,800$ {\sc loc} \lstinline|adsafe.js| library.  Various ADsafe bugs were
found in the past and this paper presents a few more
(\secref{sec:bugs}).  Note that ADsafe is a small Web sandbox
relative to larger systems like Caja.

The Caja project asked an external review team to perform a code
review~\cite{awad:cajareview}.  The findings describe many low-level details that are
similar to those we discussed above.  In addition, two higher-level
concerns stand out:
\begin{itemize}

\item ``[Caja is] hard to review. No map states invariants and points to
where they are enforced, which hurts maintainability and security.''

\item ``Documentation of TCB is necessary for reviewability and
confidence.''

\end{itemize}
These remarks identify an overarching requirement for any review: the
need for specifications so that readers can both determine whether
these fit their needs and check whether these are implemented
correctly.

\section{Verifying a Sandbox: Our Roadmap\label{sec:plan}}

\paragraph{Defining Safety}

Because humans are expensive and error-prone, and because the code review
needs to be repeated every time the program changes, it is best to
automate the review process.  However, before we begin automating anything, we need some definition of what security
means.  We focus on a definition that is specific to ADsafe, though the properties are similar to the goals of other web sandboxes.  From correspondence with ADsafe's author, we initially
obtained the following list of intended properties (rewritten slightly
to use the terminology of this paper).
\begin{adsafety}[ADsafety]\label{def:adsafety}
If the containing page does not augment built-in prototypes, and all
embedded widgets pass JSLint, then:
\begin{enumerate}
\item widgets cannot load new code at runtime, or cause \as{} to load
  new code on their behalf;
\item widgets cannot affect the \dom{} outside of their designated
subtree;
\item widgets cannot obtain direct references to \dom{} nodes; and
\item multiple widgets on the same page cannot communicate.
\end{enumerate}
\end{adsafety}

Note that the first two properties are common to sandboxes in
general---allowing arbitrary JavaScript to load at runtime compromises all
sandboxes' security goals, and all sandboxes provide mediated access to the
DOM by preventing direct access.

We also note that the assumption about built-in prototypes is often violated in
practice~\cite{mf;jw;ab:CapabilityLeaks}.  Nevertheless, like ADsafe,
we make this assumption; mitigating it is outside our scope.  
Given this definition, our goal is to produce a (mostly) automated verification that supports these properties.

\paragraph{Verifying Safety}

In this paper we perform this automation
using static types, presenting a type-based approach for defining and
verifying the invariants of ADsafe.  While one could build a custom
tool to do this, we are able to perform our verification by extending
(as discussed in section~\ref{sec:related})
a type checker~\cite{ag;cs;sk:FlowTypes} intended for
traditional type-checking of JavaScript.

We choose a static type system as our tool of choice for several
reasons.  Programmers are familiar with type systems, and ours
is mostly standard (we discuss nonstandard features in sections
\ref{sec:jslint} and \ref{sec:adsafe.js}).  This lessens the burden on
sandbox developers who need to understand what the verification is
saying about their code.  Second, our type system is much more
efficient than most whole-program
analyses or model checkers, leading to a quick procedure for checking ADsafe's runtime
library (20 seconds).  Efficency and understandability allow for
incremental use in a tight development loop.  Finally, our type system
is accompanied by a soundness proof.  This property accomplishes the
actual verification.  Thus, the features of comprehensibility,
efficiency, and soundness combine to make type checking an effective
tool for verifying some of the properties of web sandboxes.

In order to demonstrate the effectiveness of our type-based verification approach, we use type-based arguments to prove ADsafety.  We mostly
achieve this (\secref{sec:theorems}) after fixing bugs exposed by our
type checker (\secref{sec:bugs}).  The rest of this paper
presents a typed account of untrusted widgets and the ADsafe runtime.
\begin{itemize}

\item
The ADsafety claim is predicated on widgets passing the JSLint
checker.  Therefore, we need to model JSLint's restrictions.  We do
this in \secref{sec:jslint}.

\item
Once we know what we can expect from JSLint, we can verify the actual
reference monitor code in \texttt{adsafe.js} using type-checking
(\secref{sec:adsafe.js}).

\item
Before we can verify \texttt{adsafe.js}, we need to account for
the details of JavaScript source and model the browser environment in
which this code runs.  \Secref{sec:lambdajs} presents this additional work.

\end{itemize}
We discuss extensions to verify other Web sandboxes in
\secref{sec:future}.

\section{Modeling Secure Sublanguages\label{sec:jslint}}

All web sandboxes' runtime libraries expect to execute against widgets
that have been statically checked and rewritten, as shown in \figref{fig:sandbox-pic}.
These checks and rewrites enforce that widgets are written in a sublanguage of JavaScript.
This sublanguage ought to be specified explicitly.
We focus here on modeling the checks performed by JSLint, ADsafe's static checker,
which presents an interesting challenge: there is no formal specification of
the language of JavaScript programs that pass JSLint.  Instead, the
specification is implicit in the implementation of JSLint itself.
In this
section, we design a specification for JSLint-ed widgets and give
confidence in its correctness.\footnote{Because we want a strategy
  that extends to other sandboxes, we do not try to exploit the fact
  that JSLint is written in JavaScript.  The Cajoler of Caja is
  instead written in Java, and the filters and rewriters for other
  sandboxes might be written in other languages.  The strategy we
  outline here avoids both getting bogged down in the details of all
  these languages as well as over-reliance on JavaScript itself.}

Only a fraction of JSLint's static checks are related to ADsafe.  The
rest are \texttt{lint}-like code-quality checks.  JSLint also checks
the static \html{} of a widget.  Verifying this static
\html{} is beyond the scope of our work; we do not discuss it further.
We instead focus on the security-critical static JavaScript
checks in JSLint.

How is JSLint used?  The ADsafe runtime makes several assumptions
about the shape of values it receives from widgets.  These assumptions
are not documented precisely, but they correspond to various static
checks in JSLint.  To model JSLint, we reflect these checks in a
\emph{type}, called \tlint{}, which we define below.  In
\secref{sec:type-impl-corres} we discuss how this type relates to the
behavior of the JSLint implementation.

\subsection{Defining \tlint{}}

\begin{figure}[t]
  \begin{displaymath}
    \begin{array}{rcl}
      \alpha & := & \textrm{type identifiers} \\ 

      T & := & \type{Num} \mid \type{Str} \mid \type{True} \mid \type{False}
      \mid \type{Undef} \mid \type{Null} \\

      & \mid & \type{Ref}\ T \mid \forall \alpha . T \mid \mu \alpha.T \\
      & \mid & [T] T \times \ldots \times T \times T \cdots \rightarrow T \\

      & \mid & \top \mid \bot \mid T \cup T \mid T \cap T \mid $\tarray{T}$ \\

      & \mid & \{\star: F, proto: T, code: T, f: F, \ldots\}
      \\

      & \mid & (f, \ldots)^+ 
        \mid   (f, \ldots)^- \\

      F & := & T \mid \badtype \mid \tabsent{} \\


    \end{array}
  \end{displaymath}
  \caption{Type Language for \as{} and Widgets}
  \label{fig:types}
  \medskip \hrule
\end{figure}

We expect that \emph{all variables and sub-expressions} of widgets are
typable as \tlint{}.  The ADsafe runtime can thus assume that widgets
only manipulate
\tlint{}-typed values.  Our full type language is shown in
\figref{fig:types} and introduced gradually in the rest of this
section.

\paragraph{Primitives} JSLint admits JavaScript's primitive values,
with trivial types:
\begin{displaymath}
\begin{array}{rcl}
\type{Prim} & = & \type{Num}\cup\type{Str}\cup\type{True} \cup \type{False}\\
            &   & \cup \type{Null}\cup\type{Undef}
\end{array} 
\end{displaymath}
  We have separate types for \type{True} and
\type{False} because they are necessary to type-check
\lstinline|adsafe.js| (\secref{sec:adsafe.js}).  \type{Prim} is an untagged
union type, and our 
type system accounts for common JavaScript patterns for discriminating
unions.  We might initially assume that
\begin{displaymath}
\begin{array}{rcl}
\tlint & = & \type{Prim}
\end{array} 
\end{displaymath}

\paragraph{Objects and Blacklisted Fields} 

\lint{} admits object literals but blacklists  certain field
names as dangerous.  All other fields are allowed to contain widget
values.  We therefore augment the \tlint{} type to include objects.
An object type explicitly lists the names and types of various fields in an
object.  In addition, the special field $\star$ specifies the type of
all other fields:
\begin{displaymath}
  \tlint{} = 
  \begin{array}{l}
    \mu\alpha. 
    \type{Prim} \cup 
    \type{Ref}\left\{
    \begin{array}{l}
      \star : \alpha, \\
      \js{"arguments"} : \badtype, \\
      \js{"caller"} : \badtype,\\ ~ \js{"callee"} :
      \badtype, \\
      \js{"eval"} : \badtype,  \\
      \ldots \\
      \js{"toString"} : \tabsent, \\
      \js{"valueOf"} : \tabsent \\
    \end{array}\right\} \\ 
  \end{array}
\end{displaymath}

The full list of blacklisted fields is in \figref{fig:lint}.  Our
type checker signals a type error on any $\badtype$-typed field
access or assignment.  
This mirrors the behavior of JSLint, which also rejects field accesses
and assignments on blacklisted fields (e.g., \js{o["constructor"]}
is rejected by both the type checker and JSLint).

The \type{Ref} tag indicates that the object is mutable.  We
use a recursive type ($\mu$) to indicate that all other fields, $\star$,
may recursively contain \tlint{}-typed
values.\footnote{$\mu\alpha.T$ binds the type variable $\alpha$ in the
type $T$ to the whole type,
$\mu\alpha.T$. Therefore, $\alpha$ is in fact the type \tlint{}.}
\lint{} tries to ensure that objects in widgets do not have
\code{toString} and \code{valueOf} properties.  We model this with a type \tabsent{}, which ensures these fields are not present.

\tabsent{} and $\badtype$
properties are subtly different.  $\badtype$ models fields that are
intended to be inaccessible, and hence looking them up is untypable.  In contrast, the typing rule for \tabsent{} field lookup performs the lookup with the type of the \textit{proto} field, which we introduce below. \Secref{sec:type-highlights} contains the details of type-checking field access.

\paragraph{Functions} Widgets can create and apply functions, so we must
widen our \tlint{} type to admit them.  Functions in JavaScript are
objects with an internal \textit{code} field, which we add to
allowed objects:
\begin{displaymath}
\ldots\type{Ref}\left\{
\begin{array}{l}
  \codep{} : [\twindow \cup \alpha] \alpha \cdots \rightarrow \alpha, \\
  \star : \alpha, \\
  \ldots
\end{array}\right\}
\end{displaymath}
The type of the $\codep{}$ field indicates that widget-functions may
have an arbitrary number of \tlint{}-typed arguments and return
\tlint{}-typed results.\footnote{The $\alpha \cdots$ syntax is a
  literal part of the type, and means the function can be applied to
  any number of additional $\alpha$-typed arguments.}  It also
specifies that the type of the implicit \lstinline|this|-argument
(written inside brackets) may
be either \tlint{} or \twindow{}.  The type \type{Global} is not a subtype of \type{Widget}, which expresses the
underlying reason for JSLint's rejection of all widgets that contain
\lstinline|this| (see Claim 1 below).  If the \code{this}-annotation is omitted, the type
of \code{this} is $\top$.

\paragraph{Prototypes} JSLint does not allow widgets to explicitly
manipulate objects' prototypes.  However, since field lookup in
JavaScript implicitly accesses the prototypes, we specify the type of
prototypes in \tlint{}:
\begin{displaymath}
\ldots\type{Ref}\left\{
\begin{array}{l}
  proto : \type{Object} \cup \type{Function} \cup \ldots, \\
  \star : \alpha, \\
  \ldots
\end{array}\right\} \\ 
\end{displaymath}
The $\textit{proto}$ field enumerates several safe prototypes, but
notably omits DOM prototypes such as \lstinline|HTMLElement|, since
widgets should not obtain direct references to the DOM.

\paragraph{Typing Private Fields}  In addition to explicitly blacklisted
field names, JSLint also blacklists all field
names that start and end with an underscore.  This effectively
blacklists the \lstinline|__proto__| field, which gives direct access to
the prototype-chain, and the \lstinline|__nodes__| and
\lstinline|__star__| fields, which 
\lstinline|adsafe.js| uses internally to build the Bunch abstraction.  To keep our types 
simple, we enumerate these three fields instead of pattern-matching on
field names:
\begin{displaymath}
\ldots\type{Ref}\left\{
\begin{array}{l}
  \js{"___nodes___"} : $\tarray{\tdom}$ \cup \type{Undef}, \\
  \js{"__proto__"} : \badtype, \\
  \js{"___star___"} : \type{Bool} \cup \type{Undef}, \\
  \star : \alpha, \\
  \ldots
\end{array}\right\} \\ 
\end{displaymath}
The \lstinline|__proto__| field is $\badtype$-typed, like other
blacklisted fields that are never used.  However, the ADsafe runtime
uses
\lstinline|__nodes__| and \lstinline|__star__| as private fields.  The
types specify that ADsafe stores DOM references in the
\lstinline|__nodes__| field.

The full \tlint{} type in \figref{fig:lint} is a formal
specification of the shape of values that \lstinline|adsafe.js|
receives from and sends to widgets.  This type is central to our
verification of \lstinline|adsafe.js| and of JSLint.

\begin{figure}[t]
  \tlint{} = $\mu\alpha .$
  \begin{displaymath}
    \begin{array}{l}
      \type{Str} \cup \type{Num} \cup \type{Null}
      \cup \type{Bool} \cup \type{Undef}~ \cup \\
      \type{Ref} \left\{\begin{array}{l}
      proto : 
      \begin{array}{l}
        \hspace{1ex} \type{Object} \cup \type{Function} \\ 
        \cup \type {Bunch} \cup \type{Array} \cup \type{RegExp} \\
        \cup \type{String} \cup \type{Number} \cup \type{Boolean}, \\
      \end{array} \\
      \star : \alpha, \\
      code : [\twindow \cup \alpha] \alpha \cdots \rightarrow \alpha, \\
      \js{"___nodes___"} : $\tarray{\tdom}$ \cup \type{Undef}, \\
      \js{"___star___"} : \type{Bool} \cup \type{Undef}, \\
      \js{"caller"} : \badtype,~ \js{"callee"} :
      \badtype, \\
      \js{"eval"} : \badtype, \js{"prototype"} :
      \badtype,~ \\
      \js{"watch"} : \badtype, \js{"constructor"} : \badtype, \\
      \js{"__proto__"} : \badtype,~ \js{"unwatch"} :
      \badtype, \\
      \js{"arguments"} : \badtype,~ \js{"valueOf"} : \tabsent, \\
      \js{"toString"} : \tabsent \\
      \end{array}\right\} \\ 
    \end{array}
  \end{displaymath}
  \caption{The \tlint{} type}
  \label{fig:lint}
  \medskip \hrule
\end{figure}

\subsection{\tlint{} and JSLint Correspondence\label{sec:type-impl-corres}}

Though we have offered intuitive arguments for why 
\tlint{} corresponds to the checks in JSLint, we would like to
gain confidence in its correspondence with the behavior of the actual
JSLint program that sites use:
\begin{jslint}[Linted Widgets Are Typable]\label{def:lint_adequate}
If JSLint (with ADsafe checks) accepts a widget $e$, then $e$
and all of its variables and sub-expressions can be \tlint{}-typed.
\end{jslint}
We validate this claim by testing.  We use ADsafe's sample widgets as
positive tests---widgets that should be typable and lintable---and our
own suite of negative test cases (widgets that should be untypable and
unlintable).  Note the direction of the implication: an unlintable
widget may still be typable, since our type checker admits safe
widgets that \lint{} rejects.\footnote{The supplemental material
  contains examples of the differences.}
The type checker could be used as a replacement for
\lint{}'s ADsafe checks, but these tests give us confidence
that checking the \tlint{} type corresponds to what JSLint admits in
practice.

\section{Modeling JavaScript and the Browser\label{sec:lambdajs}}

Verification of a Web sandbox must account for the idiosyncrasies of
JavaScript.  It also needs to model the run-time
environment---provided by the browser---in which the sandboxed code
will execute.  Here we discuss how we model the language and the
browser.

\paragraph{JavaScript Semantics} We use the
 semantics of Guha, et al.~\cite{js_essence}, which
reduces JavaScript to a core semantics
called \ljs. This latter language models the
``essentials'' of JavaScript: prototype-based objects, first-class
functions, basic control operators, and mutation.  

\ljs\ thus omits
many of JavaScript's complexities, but it is accompanied by 
a \emph{desugaring} function that maps all
JavaScript programs (idiosyncrasies included) to behaviorally equivalent
\ljs\ programs.  The transformation explicates much of JavaScript's implicit
semantics.  Hence, we find it easier to build tools that analyze the
much smaller \ljs\ language than to directly process JavaScript.

Does desugaring faithfully map JavaScript to \ljs?  Guha, et al.\ 
test their desugaring and semantics on portions of the Mozilla 
JavaScript test
suite.  On these tests, \ljs\ programs produce exactly the same
output as JavaScript implementations.  Hence, their work
substantiates the following two claims.
\begin{desugarclaim1}[Desugaring is Total]
For all JavaScript programs $e$, $\textit{desugar}\llbracket e \rrbracket$ is defined.
\end{desugarclaim1}
\begin{desugarclaim2}[Desugar Commutes with Eval]
  For all JavaScript programs $e$, $\textit{desugar}\llbracket
  \textit{eval}_{\textit{JavaScript}}(e) \rrbracket =
  \textit{eval}_{\lambda_{\textit{JS}}}(\textit{desugar}\llbracket e\rrbracket)$.
\end{desugarclaim2}
This testing strategy, and the simplicity of implementation that \ljs\
enables, give us confidence that our tools correctly account for
JavaScript.

\begin{figure}
\begin{lstlisting}
{
  eval: $\badtype$,
  setTimeout: $(\tlint{} \rightarrow \tlint{}) \times \tlint{} \rightarrow
\type{Int}$,
  document: {
    write: $\badtype$,
    writeln: $\badtype$,
    $\ldots$
  },
  $\ldots$
}
\end{lstlisting}
\caption{A Fragment of the Type of \texttt{window}}
\label{fig:wndtyp}
\medskip \hrule
\end{figure}

\paragraph{Modeling the Browser \dom} ADsafety claims that
\lstinline|window.eval| is
not applied.  To validate this claim, we mark \lstinline|eval| with
$\badtype$ from \secref{sec:jslint}, which marks banned fields.
There are many \lstinline|eval|-like function in Web
browsers, such as \lstinline|document.write|; these are also marked
$\badtype$.  Finally, certain functions, such as
\lstinline|setTimeout|, behave like
\lstinline|eval| when given strings as arguments.  ADsafe does need to
call these functions, but it is careful to never call them with
strings.  In our type environment, we give them restrictive types that disallow
string arguments.  

\Figref{fig:wndtyp} specifies a fragment of the
type of \lstinline|window|, which carefully
specifies the type of unsafe functions in the environment.
The remaining safe \dom{} does not need to be fully specified.
\lstinline|adsafe.js| only uses a small subset of the \dom{} methods.
These methods require types.  The browser environment is therefore
modeled with $500$ lines of object types (one field per line).
This type environment is essentially
the specification of foreign \dom{} functions imported into JavaScript.

\section{Verifying the Reference Monitor\label{sec:adsafe.js}}
\lstset{language=JavaScript}

\begin{figure}
  \begin{lstlisting}
var dom = {
  append: 
  function(bunch)
  /*: $[\tlint{}\cup\twindow{}] \tlint{} \times \tlint{} \cdots \rightarrow \tlint{}$ */
    { // body of append ...  },
  combine:
  function(array)
  /*: $[\tlint{}\cup\twindow{}] \tlint{} \times \tlint{} \cdots \rightarrow \tlint{}$ */
    { // body of combine...  },
  q: 
  function (text) 
  /*: $[\tlint{}\cup\twindow{}] \tlint{} \times \tlint{} \cdots \rightarrow \tlint{}$ */
    { // body of q... },
  // ... more dom ...
};
  \end{lstlisting}
  \caption{Annotations on the \code{dom} object}
  \label{fig:annotation_example}
  \medskip\hrule
\end{figure}

In \secref{sec:jslint}, we discussed modeling the sublanguage of
widgets interacting with the sandboxing runtime.  In the case of
ADsafe and JSLint, we built up the \tlint{} type as a specification of
the kinds of values that the reference monitor, \asjs{}, can expect at
runtime.  In this section, we discuss how we use the \tlint{} type to
model the boundary between reference monitor and widget code, and
ensure that the runtime library correctly guards critical behavior.

The \tlint{} type specifies the shape of widget values that the
ADsafe runtime manipulates.  \tlint{} is therefore used pervasively in
our verification of \asjs{}.  For example, consider a
typical Bunch method:
\begin{lstlisting}
Bunch.prototype.append = function(child) {
  reject_global(this);
  var elts = child.__nodes__;
  $\ldots$
  return this;
}
\end{lstlisting}
The \type{Bunch} objects that ADsafe passes to the widget have
\lstinline{Bunch.prototype} as their \textit{proto} (see
\figref{fig:lint}), making these methods accessible.  Their use in
the widget is constrained only by JSLint, so we must type-check these methods with (only) JSLint's assumptions in mind.

For example, we might assume that the \lstinline|child| argument above should be a
\type{Bunch}, the implicit \lstinline|this| argument should also be a
\type{Bunch}, and it therefore returns a \type{Bunch}.  However,
JSLint does not provide such strong guarantees.  Consider this
example, which passes JSLint:

\begin{lstlisting}
var func = someBunch.append;
func(900, true, "junk", -7);
\end{lstlisting}

\noindent Here, \lstinline|this| is bound to \lstinline|window|,
\lstinline|child| is a number, and there are additional
arguments.  Therefore, we cannot assume that
\lstinline|append| has the type $[\type{Bunch}] \type{Bunch} \rightarrow
\type{Bunch}$.  Instead, the most precise type we can ascribe is:

\begin{displaymath}
[\tlint{} \cup \twindow{}] \tlint\cdots \rightarrow \type{Widget}
\end{displaymath}

\noindent That is, \lstinline|this| could be \type{Widget}-typed or the type of the global object, \type{Global}, and the other arguments may have any subtype of \type{Widget}, which includes strings, numbers, and other non-\type{Bunch} types.
The runtime check in \lstinline|append|'s body
(namely, \lstinline|reject_global(this)|)
is responsible for checking that
\lstinline|this| is not the global object before manipulating it.  Our
type checker recognizes such checks and narrows the broader type to
\type{Widget} after appropriate runtime checks are applied
(\secref{sec:type-highlights}).  If 
such checks were missing, the type of
\lstinline|this| would remain $\type{Widget} \cup \twindow{}$, and
\lstinline|return this| would signal a type error because
$\type{Widget} \cup \twindow{}$ is not a subtype of the stated
return type \type{Widget}.

Ascribing types to functions provided by the ADsafe runtime is
therefore trivial.  We give all the same type:
\begin{displaymath}
[\tlint{} \cup  \twindow{}] \tlint\cdots \rightarrow \tlint{}
\end{displaymath}
The type checker we extend is not ADsafe-specific, and requires
explicit type annotations.  However, since all the annotations are
identical, they are trivial to insert.
\Figref{fig:annotation_example} shows a small excerpt of such
annotations, which the checker reads from comments, so programs can
run unaltered in the browser.

\paragraph{Types for Private Functions}

\as{} also has a number of private functions, which are not exposed to
the widget.  These functions have types with capabilities the widget
does not have access to, such as \tdom{}.  For example, \as{} specifies a
\code{hunter} object, which contains functions that traverse the
\dom{} and accumulate arrays of \dom{} nodes.  These functions all
have the type $\tdom{} \rightarrow \type{Undef}$, and add to an array
\code{result} that has type \tarray{\tdom}.  \as{} can freely use
these capabilities inside the library as long as it doesn't hand them
over to the widget.  Our annotations show that it doesn't, because these
types are not compatible with \tlint.

\subsection{Type System Highlights\label{sec:type-highlights}}

In \secref{sec:jslint} and~\ref{sec:lambdajs}, we presented
types for safe objects and for values in the browser environment.  We
build upon earlier work on type systems that has been applied to
\jscript{} \cite{ag;cs;sk:FlowTypes}. In this section, we present the
non-standard portions of our type system that we use for typing
operations on objects, sensitive conditionals, and some idiosyncrasies
of \lint{} and \asjs{}.

\paragraph{Object Properties and String Set Types}

\begin{figure}
  
  \begin{mathpar}
  \inferrule[T-StringSet]
  {}
  {\Sigma; \Gamma \vdash str : (str)^+}

  \inferrule[ST-StringSet$^+$]
  {\forall f \in (f_1, \ldots), f \in (s_1, \ldots)}
  {(f_1, \ldots)^+ <: (s_1, \ldots)^+}

  \inferrule[ST-StringSet$^-$]
  {\forall f \in (f_1, \ldots), f \not \in (s_1, \ldots)}
  {(f_1, \ldots)^+ <: (s_1, \ldots)^-}
    
  \inferrule[ST-String$^+$]
  {}
  {(f_1, \ldots)^+ <: \type{Str}}

  \inferrule[ST-String$^-$]
  {}
  {(f_1, \ldots)^- <: \type{Str}}

  \inferrule[Equiv-Str]
  {}
  {\type{Str} <: ()^-}

  \end{mathpar}

  \caption{Typing and operations on string set types}
  \label{fig:stringsets}

  \medskip \hrule

\end{figure}

In \jscript, object properties (or ``fields'') are merely string
indices: even
\js{o.x} is just an alias for \js{o["x"]}.  In addition, these strings
can be computed and flow through the program before they are used to
look up fields.  Sandboxes thus deal with whitelists and blacklists of
property names.  To model this, we enrich the type language with sets
of strings.  For example, $(\js{"___nodes___"}, \js{"__proto__"})^-$ is
the type of all strings \emph{except} \js{"___nodes___"} and
\js{"__proto__"}, and $(\js{"x"}, \js{"foo"})^+$ is the type of
exactly \js{"x"} and \js{"foo"}.

\Figref{fig:stringsets} shows typing rules and operations for
string sets.  Sets support combination via unions, subtyping via
adding new strings, and subtyping of positive and negative sets.  Both
kinds of string sets can also be promoted to the common supertype of
\type{Str}, which is equivalent to the negative string set with no
entries.

\begin{figure*}
\begin{displaymath}
  \{\star\} \textrm{\ is shorthand for\ }
  \{\star: F_\star, proto: T_p, code: T_c, f_1: F_1, \ldots \}
\end{displaymath}

\begin{mathpar}
\begin{array}{lcl}
    (f_1, \ldots)^+ - (s_1, \ldots)^+ & = & \forall f_i \not \in
    (s_1, \ldots), (f_i, \ldots)^+  \\

    (f_1, \ldots)^- - (s_1, \ldots)^+ & = & (f_1, \ldots, s_1,
    \ldots)^- \\
  \end{array}

  \begin{array}{ll}
    f \in (f_1, \ldots)^+ & : \exists f_1 . f = f_1 \\
    f \in (f_1, \ldots)^- & : \forall f_1 . f \neq f_1 \\
  \end{array}
\end{mathpar}

\begin{mathpar}
  \fields_\star(\{\star\}, S) =
  \begin{array}{c}
    \left\{\begin{array}{rl}
    F_\star & : 
    \begin{array}{c}
      S_\star \neq \emptyset \textrm{\ and}\\
      F_\star \neq \tabsent{} \\
    \end{array}\\

    \bot & : \textrm{otherwise} 

    \end{array}
    \right.\\
    \textrm{where } S_\star = S - (f_1, \ldots)^+
  \end{array}
  
  \fields_p(\{\star\}, S) =
  \begin{array}{c}
    \left\{\begin{array}{rl}
    \type{Undef} & : T_p = \type{Null} \\
    fields(T_p, S_p) & : S_p \neq \emptyset \\
    \bot & : \textrm{otherwise} \\
    \end{array}\right. \\
    \textrm{where } S_p = S - (f_i \mid F_i \neq \tabsent{})^+ 
  \end{array}
\end{mathpar}

\begin{displaymath}
\begin{array}{lcl}
\fields(\{\star\}, S)
& = & 
\{T_i \mid f_i \in S \textrm{\ and\ } F_i = T_i\} \cup
\fields_\star(\{\star\}, S) \cup \fields_p(\{\star\},S) \\
\fields(T_1 \cup T_2, S) & = & \fields(T_1, S) \cup \fields(T_2, S) \\
\fields(T, \emptyset) & = & \type{$\bot$} \\
\end{array} 
\end{displaymath}

\infrule[T-Lookup]
{\Sigma; \Gamma \vdash e_o : T_o \andalso \Sigma; \Gamma \vdash e_f :
  S \andalso S <: \type{Str} \andalso T_{res} = \fields(T_o, S)}
{\Gamma \vdash \js{$e_o$[$e_f$]} : T_{res}}

\caption{Typing object lookup}
\label{fig:lookup}
\medskip\hrule
\end{figure*}

\begin{figure*}[t]
  \begin{displaymath}
    \begin{array}{|c|c|c|}
      \hline
      \textrm{Object Type $T_o$} & \textrm{String Type $S$} & \fields(T_o,S)
      \\ \hline\hline

      \{\protop: \type{Null}, \star: \type{Bool}, \code{"x"}: \type{Num}\} &
      (\code{"x"})^+ &
      \type{Num} \\
      \hline

      \{\protop: \type{Null}, \star: \type{Bool}, \code{"x"}: \type{Num}\} &
      (\code{"x"}, \code{"y"})^+ &
      \type{Num} \cup \type{Bool} \cup \type{Undef} \\
      \hline

      \{\protop: \type{Object}, \star: \type{Num}\} &
      (\code{"toString"})^+ &
      \type{Num} \cup \rightarrow \type{Str} \\
      \hline

      \{\protop: \type{Object}, \star: \type{Num}, \code{"toString"}: \tabsent{}\} &
      (\code{"toString"})^+ &
      \rightarrow \type{Str} \\
      \hline

      \{\protop: \type{Null}, \star: \type{Str}, \code{"x"}: \type{Num},
      \code{"y"}: \type{Bool}, \code{"eval"}: \badtype\} &
      (\code{"eval"})^- &
      \type{Str} \cup \type{Num} \cup \type{Bool} \cup \type{Undef} \\
      \hline

      \{\protop: \type{Null}, \star: \type{Str}, \code{"x"}: \type{Num},
      \code{"y"}: \type{Bool}, \code{"eval"}: \badtype\} &
      (\code{"eval"})^+ &
      \textrm{untypable} \\
      \hline

    \end{array}
  \end{displaymath}
  \caption{Examples of property lookup using \fields}
  \label{fig:fields_examples}
  \medskip\hrule

\end{figure*}

Equipped with string sets, we can describe the typing of object
property dereference.  When the property name is a string set, we
union the types of the properties that are members of the string set,
paying careful attention to absent fields and prototype lookup.
\Figref{fig:lookup} shows the rule \judge{T-Lookup}, with examples
shown in \figref{fig:fields_examples}.

String sets allow the type checker to avoid certain named properties,
as in the last example of \figref{fig:fields_examples}, where the
\code{"eval"} property has the bad type $\badtype$ but the string set
type of the index excludes \code{"eval"}.  The rule for property
update (not shown here) is similar but simpler, as property update in
\jscript{} does not recur inside prototypes, and only operates on the
property names of the top-level object.

\paragraph{If-Splitting} A reference monitor has
various runtime checks to ensure that protected objects---\dom{}
objects and browser functions in ADsafe's case---are only manipulated
in safe and well-defined ways.  For example, when \code{setTimeout}'s
first argument is a string, rather than a function, it exhibits
\code{eval}-like behavior, which violates \asty's constraints.  Thus we instead give
it the type
\begin{displaymath}
(\type{Widget} \rightarrow \type{Widget}) \times \type{Widget} \rightarrow \type{Num}
\end{displaymath}
Doing so forces the first argument to be a function and, in particular,
not a string.  Now consider its use:
\begin{lstlisting}
later: function (func, timeout) 
/*: $\type{Widget} \times \type{Widget} \rightarrow \type{Widget}$ */ {
  if (typeof func === "function") {
    setTimeout(func, timeout || 0);
  } else { error(); }
}
\end{lstlisting}
Because \code{ADSAFE.later} is exported to widgets, it can only assume
the \type{Widget} type for its arguments, including \code{func}.  A
traditional type checker would thus conclude that \code{func} has type
\type{Widget} everywhere in \code{later}.  Because \type{Widget} includes
\type{Str}, the invocation of \code{setTimeout} would yield a type
error---even though this is precisely what the conditional in
\code{later} is avoiding!

\emph{If-splitting} is the name for a collection of techniques that
address this problem~\cite{typed_scheme}.  Our particular solution
uses a refinement of this idea, called flow
typing~\cite{ag;cs;sk:FlowTypes}, which complements type-checking with
flow analysis.  The analysis informs the type checker that due to the
\code{typeof} check, uses of
\code{func} in the \code{then}-branch of the conditional can in fact
be \emph{refined} from the large \type{Widget} type of $\type{Str}
\cup \type{Num} \cup \ldots$ to the function type that
\code{setTimeout} requires.

\subsection{Required Refactorings\label{sec:refactorings}}

Our type system cannot type check the
ADsafe runtime as-is; we need to make some simple refactorings.
The need for these refactorings does not reflect a weakness in \as{}.
Rather, they are programming patterns that we
cannot verify with our type system.  To gain confidence that we didn't change ADsafe's behavior, we run ADsafe's sample widgets against our refactored version of
ADsafe, and they behave as expected.
We describe these refactorings below:

\paragraph{Additional \code{reject_name} Checks}
ADsafe uses \lstinline|reject_name| to check accesses and updates
to object properties in \asjs{}.  If-splitting uses these
checks to narrow string set types and type-check object property
references.  However, \as{} does not use \code{reject_name} in every
case.  For example, it uses a regular expression to parse
\dom{} queries, and uses the result to look up object properties.
   Because our type system makes conservative assumptions
about regular expressions, it would erroneously indicate that a
blacklisted field may be accessed.  Thus, we add calls to
\code{reject_name} so the type system can prove that the accesses and
 assignments are safe.

\paragraph{Inlined \code{reject_global} Checks}
Most Bunch methods start by asserting
\lstinline|reject_global(this)|, which ensures that \lstinline|this|
is \tlint{}-typed in the rest of the method.  Our type system cannot
account for such non-local side-effects, but once we inline
\lstinline|reject_global|,
if-splitting is able to refine types appropriately (for instance, in
the \lstinline|Bunch.prototype.append| example early in this section).

\paragraph{\code{makeableTagName}} 
ADsafe's whitelist of safe \dom{} elements is defined as a dictionary:
\begin{lstlisting}
var makeableTagName = 
  { "div": true, "p": true, "b": true, $\ldots$ };
\end{lstlisting}
This dictionary omits an entry for \lstinline|"script"|. The
\lstinline|document.createElement| \dom{} method creates new nodes.
We ensure that \lstinline{<script>} tags are not created by typing it as
follows:
\begin{displaymath}
\code{document.createElement} : \left(\code{"script"}\right)^{-}
\rightarrow \tdom
\end{displaymath}
ADsafe uses its tag whitelist before calling \lstinline|document.createElement|:
\begin{lstlisting}
if (makeableTagName[tagName] === true) {
  document.createElement(tagName);
}
\end{lstlisting}
Our type checker cannot account for this check.  We instead refactor
the whitelist (a trick noted
elsewhere~\cite{sm;jcm;at:WebCapabilities}):
\begin{lstlisting}
var makeableTagName = 
  { "div": "div", "p": "p", "b": "b", $\ldots$ };
\end{lstlisting}
The type of these strings are $(\code{"div"})^{+}$,
$(\code{"p"})^{+}$,$(\code{"b"})^{+}$, etc., so that
\lstinline|makeableTagName[tagName]| has type $(\code{"div"},
\code{"p"}, \code{"b"}, \ldots)^{+}$.  Since this finite set of
strings excludes \lstinline|"script"|, it now matches the argument
type of
\lstinline|createElement|.

\subsection{Cheating and Unverifiable Code\label{sec:weaknesses}}

A complex body of code like the ADsafe runtime cannot be type-checked
from scratch in one sitting.  We therefore found it convenient to
augment the type system with a \code{cheat} construct that ascribes a
given type to an expression without descending into it.  We could thus
use \code{cheat} when we encountered an uninteresting type error and
wanted to make progress.  Our goal, of course, was to ultimately
remove every \code{cheat} from the program.

We were unable to remove two \code{cheat}s, leaving eleven
unverified source lines in the 1,800 LOC ADsafe runtime.  We can, in
fact, ascribe interesting types to these functions, but checking them
is beyond the power of our type system.  The details may not be of
interest to the general reader, but the web content
contains the full body of unverified code and a discussion of its
types.

\section{ADsafety Redux\label{sec:theorems}}

Sections~\ref{sec:jslint} and~\ref{sec:adsafe.js} gave the details of
our strategy for modeling \lint{} and verifying \asjs{}.  In this
section, we combine these results and relate it to the original
definition of \asty{} (\defref{def:adsafety}).  The use of a
type system allows us to make straightforward, type-based arguments of
safety for the components of \as{}.

The lemmas below formally reason about type-checked widgets. 
\Claimref{def:lint_adequate} (\secref{sec:type-impl-corres})
establishes that linted widgets are in fact typable.  Therefore,
\emph{we do not need to type-check widgets}.  Widget programmers can
continue to use JSLint and do not need to know about our type checker.
However, given the benefits of uniformity provided by a type checker
over ad hoc methods like JSLint (\secref{sec:bugs} details one exploit that
resulted from such an ad hoc approach), programmers may be well served to use
our type checker instead.

\paragraph{Type Soundness}

Most type systems come with a soundness theorem that is stated as
\emph{progress} (well-typed programs do not error) and
\emph{preservation} (well-typed programs do not violate their types).

We do not attempt to establish progress.
Establishing it would require many more refactorings in the ADsafe
runtime, and many lintable widgets would be untypable.  Because
runtime errors are perfectly
acceptable (they halt execution before something bad happens),
we relax some of the typing rules in 
an existing type system~\cite{ag;cs;sk:FlowTypes}---which does exhibit
progress---to instead allow some
JavaScript errors (e.g., applying non-function values or looking up
fields of \js{null}).  
We do still need an ``untyped progress'' theorem that states that our
JavaScript semantics fully models all error cases.  This  theorem is
provided by Guha, et al.~\cite{js_essence}.

We restate and prove preservation for the extensions to Guha et al.'s type system, which is applicable to \emph{all} JavaScript
programs.\footnote{For the formal proof, see Guha et
al.~\cite{ag;cs;sk:FlowTypes} and the supplemental material on the web.}
Stated formally:

\begin{preservation}[Type Preservation]
  If, for an expression $e$, type $T$, environment $\Gamma$ and
  abstract heap $\Sigma$, 

  \begin{enumerate}
  \item $\Sigma \vdash \sigma$,
  \item  $\Sigma; \Gamma \vdash e : T$, and
  \item  $\sigma e \rightarrow \sigma' e'$;
  \end{enumerate}

\noindent then there exists a $\Sigma'$ with $\Sigma' \vdash \sigma'$ and
$\Sigma'; \Gamma \vdash e' : T$.
\end{preservation}

\noindent Our assumed environment (\secref{sec:lambdajs}) provides the abstract heap
$\Sigma$ and abstract environment $\Gamma$, which
model the initial state of the browser, $\sigma$.  
Given this
lemma, we can make type-based statements about the combination of widgets and
\asjs:

\newtheorem{adsafetythm}{Theorem}

\begin{adsafetythm}[ADsafety]
For all widgets $p$, if

\begin{enumerate}

\item all subexpressions of $p$ are \tlint{}-typable,

\item \asjs{} is typable,

\item \asjs{} runs before $p$, and

\item $\sigma p \rightarrow \sigma' p'$ (single-step reduction),
\end{enumerate}

\noindent then at every step $p'$, $p'$ also has the type \tlint{}.
\end{adsafetythm}
This theorem says that for all widgets $p$ whose subexpressions are
\tlint{}-typed, if \asjs{} type-checks and runs in the browser
environment, $p$ can take any number of steps and still have the
\tlint{} type.  Since types are preserved, two further key lemmas hold during execution:

\begin{adsafe_no_eval}[Widgets cannot load new code at runtime] 
For all widgets $e$, if all variables and sub-expressions of $e$ are \tlint{}-typed, then
$e$ does not load new code.
\end{adsafe_no_eval}
By \secref{sec:lambdajs}, \lstinline|eval|-like functions are
$\badtype$-typed, hence cannot be referenced by widgets or by the ADsafe
runtime.  Furthermore,
functions that only \lstinline|eval| when given strings, such as
\lstinline|setTimeout|, have restricted types that disallow
\lstinline|string|-typed arguments.  Therefore, neither the widget nor
the ADsafe runtime can load new code. \qed

\begin{adsafe_no_dom}[Widgets do not obtain DOM references]
\label{def:lint_no_eval}For all widgets $e$, if all variables and sub-expressions of $e$ are
\tlint{}-typed, then $e$ does not obtain direct \dom{} references.
\end{adsafe_no_dom}
The type of \dom{} objects is not subsumed by the \tlint{} type.
All functions in the ADsafe runtime have the
type:
\[ [\tlint{} \cup
  \twindow{}] \tlint{}\cdots \rightarrow \tlint{} \]
Thus, functions in the ADsafe runtime do not leak \dom{} references, as
long as they are only applied to \tlint{}-typed values.
Since all subexpressions of the widget $e$ are \tlint{}-typed, all
values that $e$ passes to the ADsafe runtime are \tlint{}-typed.
By the same argument, $e$ cannot directly manipulate \dom{} references
either. \qed

\paragraph{Widgets can only manipulate their DOM subtree}

We cannot prove this claim with our tools.  JSLint enforces this
property by also verifying the static \html{} of widgets; it ensures that
all element {\sc id}s are prefixed with the widget's {\sc id}.  The wrapper
for \lstinline|document.getElementById| ensures that the widget {\sc id} is a
prefix of the element {\sc id}.  Verifying JSLint's \html{} checks is beyond the
scope of this work.

In addition, the wrapper for \lstinline|Element.parentNode| checks to
see if the current element is the root of the widget's \dom{} subtree.  It
is not clear if our type checker can express this property without
further extensions.

\paragraph{Widgets cannot communicate}

This claim is false; \secref{sec:stylebug} presents a
counterexample.

\section{Bugs Found in ADsafe\label{sec:bugs}}
\lstset{language=JavaScript}

We have implemented the type system presented in
this paper, and applied it to the ADsafe source.  The implementation
is about 3,000 LOC, and takes 20 seconds to check \asjs{}
(mainly due to the presence of recursive types).  In some cases,
type-checking failed due to the weakness of the type checker; these
issues are discussed in \secref{sec:refactorings}.  The other
failures, however, represent genuine errors in ADsafe that were
present in the production system.  The same applies to instances where
JSLint and our typed model of it failed to conform.  All the errors
listed below have been reported, acknowledged by the author, and
fixed.

\paragraph{Missing Static Checks\label{sec:staticbugs}}

\begin{figure}
\begin{lstlisting}
ADSAFE.go("AD_", function (dom, lib) { 
  var myWindow, fakeNode, fakeBunch, realBunch;

  fakeNode =  {
    appendChild: function(elt) { 
       myWindow = elt.ownerDocument.defaultView;
    }, 
    tagName: "div",
    value: null
  };

  fakeBunch = {"___nodes___": [fakeNode]};
  
  realBunch = dom.tag("p");
  fakeBunch.value = realBunch.value;
  fakeBunch.value(""); // calls phony appendChild
    
  myWindow.alert("hacked");
});
\end{lstlisting}

\caption{Exploiting JSLint}
\label{fig:fakebunch}
\medskip \hrule
\end{figure}

JSLint inadvertently allowed widgets to include underscores in quoted field
names.  In particular, the following expression was deemed safe:
\begin{lstlisting}
fakeBunch = { "__nodes__": [ fakeNode ] };
\end{lstlisting}
A malicious widget could then create an object with an \lstinline|appendChild|
method, and trick the \as{} runtime into invoking it with a direct
reference to an \html{} element, which is enough to obtain
\lstinline|window| and violate ADsafety:
\begin{lstlisting}
fakeNode = {
  appendChild: function(elt) { 
    myWindow = elt.ownerDocument.defaultView;
  }
};
\end{lstlisting}
The full exploit is in \figref{fig:fakebunch}.

This bug manifested as a discrepancy between our model of
JSLint as a type checker and the real JSLint.  Recall from
\secref{sec:jslint}
that all expressions in widgets must have type \tlint{} (defined in
\figref{fig:lint}). For
\lstinline|{ "__nodes__": [fakeNode] }| to type as \tlint{}, the
\lstinline|"__nodes__"| field must have type
\tarray{\tdom}$ \cup \type{Undef}$.  However, \lstinline|[fakeNode]| has
type \tlint{}, which signals the error.

JSLint similarly allowed \lstinline|"__proto__"| and other fields to
appear in widgets.  We did not investigate whether they can be
exploited as above, but setting them causes unanticipated behavior.
Fixing JSLint was simple once our type checker found the error.  (An
alternative solution would be to use our type system as a replacement
for JSLint.)  We note that when the ADsafe option of
JSLint was first
announced,\footnote{\url{tech.groups.yahoo.com/group/caplet/message/44}}
its author offered:
\begin{quote}
If [a malicious client] produces no errors when linted with
the ADsafe option, then I will buy you a plate of shrimp.
\end{quote}
After this error report, he confirmed, ``I do believe that I owe you a
plate of shrimp''.

\paragraph{Missing Runtime Checks}

\begin{figure}
\begin{lstlisting}
ADSAFE.go("AD_", function (dom, lib) { 
  var called = false;
  var obj = {
    "toString": function() { 
      if (called) { 
        return "url(evil.xml#exp)"; 
      }
      else {
        called = true;
        return "dummy";
      }
    }
  };
  dom.append(dom.tag("div"));
  dom.q("div").style("MozBinding", o);
});

<!-- evil.xml -->
<?xml version="1.0"?>
<bindings><binding id="exp">
<implementation><constructor>
document.write("hacked")
</constructor></implementation>
</binding></bindings>          
\end{lstlisting}
\caption{Firefox-specific Exploit for ADsafe}
\label{f:xbl}
\medskip \hrule
\end{figure}

Many functions in \asjs{} incorrectly assumed that they
were applied to primitive strings.  For example,
\lstinline|Bunch.prototype.style| began with the following check, to
ensure that widgets do not programmatically load external resources via
\css{}:
\begin{lstlisting}
Bunch.prototype.style = function(name, value) {
  if (/url/i.test(value)) { // regex match?
    error();
  }
  ...
};
\end{lstlisting}
Thus, the following widget code would signal an error:
\begin{lstlisting}
someBunch.style("background", 
  "url(http://evil.com/image.jpg)");
\end{lstlisting}
The bug is that if \lstinline|value| is an object instead of a string,
the regular-expression \lstinline|test| method will invoke
\lstinline|value.toString()|.

A malicious widget can construct an object with a stateful
\lstinline|toString| method that passes the test when first applied, and
subsequently returns a malicious {\sc url}. In Firefox, we can use such an
object to load an {\sc xbl}
resource\footnote{\url{https://developer.mozilla.org/en/XBL}} that
contains arbitrary JavaScript (\figref{f:xbl}).

We ascribe types to JavaScript's built-ins to prevent implicit type
conversions.  Therefore, we require the argument of
\lstinline|Regexp.test| to have type $\type{Str}$.  However, since
\lstinline|Bunch.prototype.style| can be invoked by widgets, its type
is $\tlint{} \times \tlint{} \rightarrow \tlint{}$, and thus the type
of \lstinline|value| is \tlint{}.

This bug was fixed by adding a new \lstinline|string_check| function
to ADsafe, which is now called in $18$ functions.  All these functions
are not otherwise exploitable, but a missing check would cause
unexpected behavior. The fixed code is typable.

\paragraph{Counterexamples to Non-Interference\label{sec:stylebug}}

Finally, a type error in \lstinline|Bunch.prototype.getStyle| helped
us generate a counterexample to ADsafe's claim of widget
noninterference (\defref{def:adsafety}, part 4). The
\lstinline|getStyle| method is available to widgets, so its type must be
$\tlint{} \rightarrow \tlint{}$.  The following code is the essence of
\lstinline|getStyle|:
\begin{lstlisting}
Bunch.prototype.getStyle = function (name) {
  var sty;
  reject_global(this);
  sty = window.getComputedStyle(this.__node__);
  return sty[name];
}
\end{lstlisting}
The bug above is that \lstinline|name| is unchecked, so it may index
arbitrary fields, such as \lstinline|__proto__|:
\begin{lstlisting}
someBunch.getStyle("__proto__");
\end{lstlisting}
This gives the widget a reference to the prototype of the browser's
\lstinline|CSSStyleDeclaration| objects.  Thus the return type of the
body is not $\tlint{}$, yielding a type error.

A widget cannot exploit this bug in isolation.  However, it can
replace built-in methods of \css{} style objects and interfere with the
operation of the hosting page and other widgets that manipulate styles
in JavaScript.

This bug was fixed by adding a \lstinline|reject_name| check that is
now used in this and other methods.  Despite the fix, ADsafe still
cannot enforce non-interference, since widgets can reference and
affect properties of other shared built-ins:
\begin{lstlisting}
var arr = [ ];
arr.concat.channel = "shared data";
\end{lstlisting}
The author of ADsafe pointed out the above example and retracted the
claim of non-interference.

\paragraph{Prior Exploits}

Before and during our implementation, other exploits were found in
\as{} and reported~\cite{sm;jcm;at;FiltersRewritingWrappers,sm;jcm;at:EnforceJavaScriptSubsets,sm;jcm;at:WebCapabilities}. We
have run our type checker on the exploitable code, and
our tools catch the bugs and report type errors.

\paragraph{Fixing Bugs and Tolerating Changes}

Each of our bug reports resulted in several changes to the source,
which we tracked.  In addition to these changes, \asjs{} also
underwent non-security related refactorings during the course of this
work.  Despite not providing its author our type checker, we were
easily able to continue type-checking the code after these changes.
One change involved adding a number of new \code{Bunch} methods to
extend the {\sc api}.  Keeping up-to-date was a simple task, since all the
new \code{Bunch} methods could be quickly annotated with the \tlint{}
type and checked.  In short, our type checker has shown robustness in
the face of program edits.

\section{Beyond ADsafe\label{sec:future}}
\lstset{language=JavaScript}

Our security type system is capable of verifying useful properties
about JavaScript programs in general.  Sections \ref{sec:jslint},
\ref{sec:lambdajs}, and \ref{sec:adsafe.js} present carefully
crafted \emph{types} that we ascribe to the browser API and \asjs{}, and
use to model widget programs.  Proving these types hold over the
ADsafe runtime library and JSLint-ed widgets guarantees robust
sandboxing properties for ADsafe.

Verifications for other sandboxes would require the design of new
\emph{types}, to accurately model checked, rewritten programs and
their interface to the sandbox, but not necessarily a new
\emph{type system}.  Indeed, 
our type-based strategy provides a concrete roadmap for sandbox
designers:

\begin{enumerate}

\item Formally specify the language of widgets using a type system;

\item use this specification to define the interface between 
the sandbox and untrusted code; and,

\item check that the body of the sandbox adheres to this interface by
type-checking. 
\end{enumerate}

In particular, developers of \emph{new} sandboxes should be aware of this
strategy.  Rather than trying to retrofit the type system's features onto
existing static checks, the sandbox designer can work with the type system to
guarantee safety constructively from the start.
Tweaks and extensions to the type system are
certainly possible---for example, one may want to design a sandboxing
framework that forbids applying non-function values and looking up
fields of \js{null}, which the current type system
allows~(\secref{sec:theorems}).

ADsafe shares many programming patterns with other Web sandboxes
(\secref{sec:codereview}), but doesn't cover the full range of their
features.  We outline some of the extensions that could be used to
verify them here:

\paragraph{Reasoning About Strings} Our type system lets programmers
reason about finite sets of strings and use these sets to lookup fields
in objects.  To verify Caja, we would need to reason about string
patterns.  For example, Caja uses the field named \lstinline|"foo" + "_w__"|
to store a flag that determines if the field \lstinline|"foo"| is
writable.

\paragraph{Abstracting Runtime Tests}
Our type system accounts for inlined runtime checks, but requires some
refactorings when these checks are abstracted into predicates.  Larger
sandboxes, like Caja, have more predicates, so refactoring them all
would be infeasible.  We could instead use ideas from occurrence
typing~\cite{typed_scheme}, which accounts for user-defined
predicates.

\paragraph{Modeling the Browser Environment}
ADsafe wraps a small subset of the \dom{} {\sc api} and we manually check that
this subset is appropriately typed in the initial type environment.
This approach does not scale to a sandbox that wraps more of the \dom.
If the type environment were instead derived from the C++ \dom{}
implementation, we would have significantly greater confidence in our
environmental assumptions.

\section{Related Work\label{sec:related}}

\paragraph{Verifying JavaScript Web Sandboxes} ADsafe~\cite{adsafe},
BrowserShield~\cite{cr;jd;hjw;od;se:BrowserShield}, Caja~\cite{caja},
and \fbjs~\cite{fbjs} are archetypal Web sandboxes that use static and
dynamic checks to safely host untrusted widgets.  However, the semantics
of JavaScript and the browser environment conspire to make JavaScript
sandboxing difficult~\cite{js_essence,sm;jm;at:JavaScriptSemantics}.

Maffeis et al.~\cite{sm;jcm;at;FiltersRewritingWrappers} use their
JavaScript semantics to develop a miniature sandboxing system and prove it
correct.  Armed with the insight gained by their semantics and proofs,
they find bugs in \fbjs{} and \as{} (which we also catch).  However, they do not mechanically
verify the
JavaScript code in these sandboxes.  They also formalize capability
safety and prove that a Caja-like subset is capability
safe~\cite{maffeis:caps}.  However, they do not verify the Caja
runtime or the actual Caja subset.  In contrast, we verify the source
code of the ADsafe runtime and account for ADsafe's static checks.

Taly, et al.~\cite{taly:cfa} develop a flow analysis to find bugs in
the ADsafe runtime (that we also catch).  They simplify the analysis by modeling
ECMAScript 5 strict mode, which is not fully implemented in any
current Web browser.  In contrast, ADsafe is designed to run on current browsers,
and thus supports older and more permissive versions of JavaScript.
We use the semantics and tools of Guha, et al.~\cite{js_essence},
which does not limit itself to the strict mode, so we find new bugs in the
ADsafe runtime.  In addition, Taly, et al.~use a simplified model
of JSLint.  In contrast, we provide a detailed, type-theoretic account
of JSLint, and also test it.  We can thus find security bugs in JSLint as
well.

Lightweight Self-Protecting
JavaScript~\cite{jm;php;ds:SafeWrappersSanePolicies,phung:spjs} is a
unique sandbox that does not transform or validate widgets.  It
instead solely uses reference monitors to wrap capabilities.  These
are modeled as security automata, but the model ignores the semantics
of JavaScript.  In contrast, this paper and the aforementioned works
are founded on detailed JavaScript semantics.

Yu, et al.~\cite{yu:corescript} use JavaScript sandboxing techniques to
enforce various security policies on untrusted code.  Their semantic
model, CoreScript, simplifies the \dom{} and scripting language.
CoreScript cannot be used to mechanically verify the JavaScript
implementation of a Web sandbox, which is what we present in this paper.

\paragraph{Modeling the Web Browser}  
There are formal models of Web browsers that are tailored to model
whole-browser security
properties~\cite{da;ab;pel;jm;ds:FormalWebFoundation,ab;bcp:FeatherweightFirefox}.
These do not model JavaScript's semantics in any detail and are
therefore orthogonal to semantic models of
JavaScript~\cite{js_essence,sm;jm;at:JavaScriptSemantics} that are
used to reason about language-based Web sandboxes.  In particular,
ADsafe's stated security goals are limited to statements about
JavaScript and the \dom{} (\secref{sec:plan}).  Therefore, we do not
require a comprehensive Web-browser model.

\paragraph{Static Analysis of JavaScript} GateKeeper~\cite{sg;bj:GATEKEEPER} uses a
combination of program analysis and runtime checks to apply and verify
security policies on JavaScript widgets.  GateKeeper's program analysis is designed to model more complex properties of untrusted code than we address by modeling JSLint.
However, the soundness of
its static analysis is proven relative to only a restricted
sub-language of JavaScript, whereas \ljs\ handles the full language.
In addition, they do not demonstrate the validity of their run-time
checks.

Chugh et al.~\cite{chugh:sif} and {\sc vex}~\cite{bandhakavi:vex} use program analysis to detect
possibly malicious information flows in JavaScript.  Our type system
cannot specify information flows, although we do use it to discover that
ADsafe fails to enforce a desirable information flow property.  
{\sc vex}'s authors acknowledge that it is unsound, and Chugh et al. do not provide a proof of soundness for their flow analysis.  Our type system and analysis are proven sound.

Other static analyses for
JavaScript~\cite{guha:ovid,jensen:tajs,jensen:lazy} are not
specifically designed to encode and check security.

\paragraph{Type Systems} Our type checker is based on that 
of Guha, et al.~\cite{ag;cs;sk:FlowTypes}.  Theirs has a
restrictive type system for objects that we fully replace to type
check ADsafe.  We also add simple extensions to their
\emph{flow typing} system to account for additional kinds of runtime
checks employed by ADsafe.  Their paper surveys other JavaScript type
systems~\cite{anderson:inference,heidegger:recency} that can
type-check other patterns but have not been used to verify
security-critical code, which is the goal of this paper.  Our
treatment of objects is also derived from
\textsc{ML-ART}~\cite{dr:ML-ART}, but accounts for JavaScript 
features and patterns such as function objects, prototypes, and
objects as dictionaries.

\paragraph{Language-Based Security} Schneider et
al.~\cite{bfs;gm;rh:LanguageBasedSecurity} survey the design and
type-based verification of language-based security systems.
JavaScript Web sandboxes are inlined reference
monitors~\cite{erlingsson:thesis}.  Guha, et al.~\cite{js_essence}
offer a type-based strategy to verify these, but their
approach---which depends on building a custom type rule around each
check in the reference monitor---does not scale to a program of the
size of ADsafe.  Furthermore, their custom rules essentially hand-code
if-splitting, which we obtain directly from the underlying type
system.

Cappos, et al.~\cite{jc;ad;jr;js;ib;cb;ak;ta:SandboxContainment} present
a layered approach to building language sandboxes that prevents bugs in
higher layers from breaking the abstractions and assurances provided by
lower layers.  They use this approach to build a new sandbox for Python,
whereas we verify an existing, third-party JavaScript sandbox.  However,
our verification techniques could easily be used from the onset to build
a new sandbox that is secure by construction.

\paragraph{IFrames}
IFrames are widely used for widget isolation.  However, JavaScript
that runs in an IFrame can still open windows, communicate with
servers, and perform other operations that a Web sandbox disallows.
Furthermore, inter-frame communication is difficult when desired;
there are proposals to enhance IFrames to make communication easier
and more secure~\cite{cj;hw:Subspace}.  Language-based sandboxing is
somewhat orthogonal in scope, is more flexible, and does not require
changes to browsers.

\paragraph{Runtime Security Analysis of JavaScript}

There are various means to secure widgets that do not employ
language-based security.  Some systems rely on modified browsers,
additional client software, or proxy
servers~\cite{dewald:ADSandbox,dhawan:extinfflows,jim:beep,kiciman:ajaxscope,louw:AdJail,meyerovich:ConScript,yu:corescript}.  Some of these propose alternative Web programming
APIs that are designed to be secure.  Language-based sandboxing has
the advantage of working with today's browsers and deployment methods,
but our verification ideas could potentially apply to the design of
some of these systems, too.

\subsection*{Acknowledgments}

We thank Douglas Crockford for discussions, open-mindedness, and
insightful feedback (and the promise of certain crustaceans); Mark
S.~Miller for enlightening discussions; Matthias Felleisen, Andrew
Ferguson, and David Wagner for numerous helpful comments that helped
us understand weaknesses in exposition; the NSF for financial support;
and StackOverflow, as well as Claudiu Saftoiu (our lower-latency
version of StackOverflow), for unflagging attention to detail.

\bibliographystyle{abbrv}
\bibliography{bibtex,arjun}

\begin{thebibliography}{10}

\bibitem{da;ab;pel;jm;ds:FormalWebFoundation}
D.~Akhawe, A.~Barth, P.~E. Lam, J.~Mitchell, and D.~Song.
\newblock {Towards a Formal Foundation of Web Security}.
\newblock In {\em {IEEE Computer Security Foundations Symposium}}, 2010.

\bibitem{anderson:inference}
C.~Anderson, P.~Giannini, and S.~Drossopoulou.
\newblock Towards type inference for {JavaScript}.
\newblock In {\em {European Conference on Object-Oriented Programming}}, 2005.

\bibitem{ja:ReferenceMonitor}
J.~P. Anderson.
\newblock {Computer Security Technology Planning Study}.
\newblock Technical Report ESD-TR-73-51, {Deputy for Command and Management
  Systems, HQ Electronic Systems Division (AFSC)}, {L. G. Handscom Field,
  Bedford, Massachusetts 01730}, October 1972.

\bibitem{awad:cajareview}
I.~Awad, T.~Close, A.~Felt, C.~Jackson, B.~Laurie, F.~Lee, K.-P. Lee, D.-S.
  Hopwood, J.~Nagra, E.~Sachs, M.~Samuel, M.~Stay, and D.~Wagner.
\newblock {Caja} external security review.
\newblock Technical report, Google Inc., 2008.
\newblock
  \url{http://google-caja.googlecode.com/files/Caja_External_Security_Review_v2.pdf}.

\bibitem{bandhakavi:vex}
S.~Bandhakavi, S.~T. King, P.~Madhusudan, and M.~Winslett.
\newblock {VEX}: Vetting browser extensions for security vulnerabilities.
\newblock 2010.

\bibitem{ab;bcp:FeatherweightFirefox}
A.~Bohannon and B.~C. Pierce.
\newblock {Featherweight Firefox: Formalizing the Core of a Web Browser}.
\newblock In {\em {Usenix Conference on Web Application Development
  (WebApps)}}, 2010.

\bibitem{jc;ad;jr;js;ib;cb;ak;ta:SandboxContainment}
J.~Cappos, A.~Dadgar, J.~Rasley, J.~Samuel, I.~Beschastnikh, C.~Barsan,
  A.~Krishnamurthy, and T.~Anderson.
\newblock {Retaining Sandbox Containment Despite Bugs in Privileged Memory-Safe
  Code}.
\newblock In {\em {ACM Conference on Computer and Communications Security
  (CCS)}}, 2010.

\bibitem{chugh:sif}
R.~Chugh, J.~A. Meister, R.~Jhala, and S.~Lerner.
\newblock Staged information flow for {JavaScript}.
\newblock 2009.

\bibitem{adsafe}
D.~Crockford.
\newblock {ADSafe}.
\newblock \url{www.adsafe.org}, 2011.

\bibitem{dewald:ADSandbox}
A.~Dewald, T.~Holz, and F.~C. Freiling.
\newblock {ADSandbox: Sanboxing JavaScript to fight Malicious Websites}.
\newblock In {\em {Symposium On Applied Computing (SAC)}}, 2010.

\bibitem{dhawan:extinfflows}
M.~Dhawan and V.~Ganapathy.
\newblock Analyzing information flow in {JavaScript}-based browser extensions.
\newblock In {\em Computer Security Applications Conference}, 2009.

\bibitem{erlingsson:thesis}
{\'U}.~Erlingsson.
\newblock {\em The Inlined Reference Monitor Approach to Security Policy
  Enforcement}.
\newblock PhD thesis, Cornell University, 2003.

\bibitem{fbjs}
{Facebook}.
\newblock {FBJS}, 2011.
\newblock \url{http://developers.facebook.com/docs/fbjs/}.

\bibitem{mf;jw;ab:CapabilityLeaks}
M.~Finifter, J.~Weinberger, and A.~Barth.
\newblock {Preventing Capability Leaks in Secure JavaScript Subsets}.
\newblock In {\em {Network and Distributed System Security Symposium}}, 2010.

\bibitem{sg;bj:GATEKEEPER}
S.~Guarnieri and B.~Livshits.
\newblock {GATEKEEPER: Mostly static enforcement of security and reliability
  policies for JavaScript code}.
\newblock In {\em {USENIX Security Symposium (SSYM)}}, {2009}.

\bibitem{guha:ovid}
A.~Guha, S.~Krishnamurthi, and T.~Jim.
\newblock Using static analysis for {Ajax} intrusion detection.
\newblock In {\em International World Wide Web Conference}, 2009.

\bibitem{js_essence}
A.~Guha, C.~Saftoiu, and S.~Krishnamurthi.
\newblock {The Essence of JavaScript}.
\newblock In {\em {European Conference on Object-Oriented Programming}}, 2010.

\bibitem{ag;cs;sk:FlowTypes}
A.~Guha, C.~Saftoiu, and S.~Krishnamurthi.
\newblock {Typing Local Control and State Using Flow Analysis}.
\newblock In {\em {European Symposium on Programming}}, 2011.

\bibitem{heidegger:recency}
P.~Heidegger and P.~Thiemann.
\newblock Recency types for dynamically-typed, object-based languages: Strong
  updates for {JavaScript}.
\newblock In {\em {ACM SIGPLAN International Workshop on Foundations of
  Object-Oriented Languages}}, 2009.

\bibitem{cj;hw:Subspace}
C.~Jackson and H.~J. Wang.
\newblock {Subspace: Secure Cross-Domain Communication for Web Mashups}.
\newblock In {\em International World Wide Web Conference}, 2007.

\bibitem{jensen:tajs}
S.~H. Jensen, A.~M{\o}ller, and P.~Thiemann.
\newblock Type analysis for {JavaScript}.
\newblock In {\em International Static Analysis Symposium}, 2009.

\bibitem{jensen:lazy}
S.~H. Jensen, A.~M{\o}ller, and P.~Thiemann.
\newblock Interprocedural analysis with lazy propagation.
\newblock In {\em International Static Analysis Symposium}, 2010.

\bibitem{jim:beep}
T.~Jim, N.~Swamy, and M.~Hicks.
\newblock {BEEP}: Browser-enforced embedded policies.
\newblock In {\em International World Wide Web Conference}, 2007.

\bibitem{kiciman:ajaxscope}
E.~K{\i}c{\i}man and B.~Livshits.
\newblock {AjaxScope}: A platform for remotely monitoring the client-side
  behavior of web 2.0 applications.
\newblock 2007.

\bibitem{louw:AdJail}
M.~T. Louw, K.~T. Ganesh, and V.~Venkatakrishnan.
\newblock {AdJail: Practical enforcement of confidentiality and integrity
  policies on Web advertisements}.
\newblock In {\em {USENIX Security Symposium (SSYM)}}, 2010.

\bibitem{sm;jm;at:JavaScriptSemantics}
S.~Maffeis, J.~Mitchell, and A.~Taly.
\newblock {An Operational Semantics for JavaScript}.
\newblock In {\em ASIAN Symposium on Programming Languages and Systems}, pages
  307--325, 2008.

\bibitem{sm;jcm;at;FiltersRewritingWrappers}
S.~Maffeis, J.~C. Mitchell, and A.~Taly.
\newblock {Isolating JavaScript with Filters, Rewriting, and Wrappers}.
\newblock In {\em {European Symposium on Research in Computer Security
  (ESORICS)}}, 2009.

\bibitem{sm;jcm;at:EnforceJavaScriptSubsets}
S.~Maffeis, J.~C. Mitchell, and A.~Taly.
\newblock Run-time enforcement of secure javascript subsets.
\newblock In {\em {W2SP}'09}. {IEEE}, 2009.

\bibitem{sm;jcm;at:WebCapabilities}
S.~Maffeis, J.~C. Mitchell, and A.~Taly.
\newblock {Object Capabilities and Isolation of Untrusted Web Applications}.
\newblock In {\em IEEE Symposium on Security and Privacy}. {IEEE}, 2010.

\bibitem{maffeis:caps}
S.~Maffeis, J.~C. Mitchell, and A.~Taly.
\newblock Object capabilities and isolation of untrusted {Web} applications.
\newblock 2010.

\bibitem{jm;php;ds:SafeWrappersSanePolicies}
J.~Magazinius, P.~H. Phung, and D.~Sands.
\newblock {Safe Wrappers and Sane Policies for Self Protecting JavaScript}.
\newblock In {\em {OWASP AppSec Research}}, 2010.

\bibitem{meyerovich:ConScript}
L.~Meyerovich and B.~Livshits.
\newblock Conscript: Specifying and enforcing fine-grained security policies
  for javascript in the browser.
\newblock In {\em IEEE Symposium on Security and Privacy}, 2010.

\bibitem{caja}
M.~S. Miller, M.~Samuel, B.~Laurie, I.~Awad, and M.~Stay.
\newblock {Caja: Safe active content in sanitized JavaScript}.
\newblock Technical report, Google Inc., 2008.
\newblock {\tt\small
  http://google\-caja.googlecode.com/files/caja-spec\--2008\--06\--07.pdf}.

\bibitem{phung:spjs}
P.~H. Phung, D.~Sands, and A.~Chudnov.
\newblock Lightweight self-protecting {JavaScript}.
\newblock 2009.

\bibitem{cr;jd;hjw;od;se:BrowserShield}
C.~Reis, J.~Dunagan, H.~J. Wang, O.~Dubrovsky, and S.~Esmeir.
\newblock {BrowserShield: Vulnerability-Driven Filtering of Dynamic HTML}.
\newblock In {\em {Symposium on Operating Systems Design and Implementation}},
  2006.

\bibitem{dr:ML-ART}
D.~R\'emy.
\newblock {Programming objects with ML-ART, an extension to ML with abstract
  and record types}.
\newblock In M.~Hagiya and J.~Mitchell, editors, {\em Theoretical Aspects of
  Computer Software}, volume 789 of {\em {Springer Lecture Notes in Computer
  Science}}, pages 321--346. Springer Berlin / Heidelberg, 1994.

\bibitem{bfs;gm;rh:LanguageBasedSecurity}
F.~B. Schneider, G.~Morrisett, and R.~Harper.
\newblock {A Language-Based Approach to Security}.
\newblock In R.~Wilhelm, editor, {\em Informatics}, volume 2000 of {\em
  {Springer Lecture Notes in Computer Science}}, pages 86--101. Springer Berlin
  / Heidelberg, 2001.

\bibitem{taly:cfa}
A.~Taly, {\'U}.~Erlingsson, M.~S. Miller, J.~C. Mitchell, and J.~Nagra.
\newblock Automated analysis of security-critical {JavaScript} {APIs}.
\newblock 2011.

\bibitem{typed_scheme}
S.~Tobin-Hochstadt and M.~Felleisen.
\newblock {The Design and Implementation of Typed Scheme}.
\newblock In {\em {ACM SIGPLAN-SIGACT Symposium on Principles of Programming
  Languages (POPL)}}, pages 395--406, 2008.

\bibitem{yu:corescript}
D.~Yu, A.~Chander, N.~Islam, and I.~Serikov.
\newblock Javascript instrumentation for browser security.
\newblock In {\em {ACM SIGPLAN-SIGACT Symposium on Principles of Programming
  Languages (POPL)}}, 2007.

\bibitem{cy;hw:InsecureJSPractices}
C.~Yue and H.~Wang.
\newblock {Characterizing Insecure JavaScript Practices on the Web}.
\newblock In {\em International World Wide Web Conference}, 2009.

\end{thebibliography}

\appendix

\section{Typing and JSLint: An Example\label{sec:jslint-vs-type}}

Consider the following malicious widget:
\begin{lstlisting}
var fakeBunch = { 
  __nodes__ : [ {
    appendChild: function(elt) {
      myWindow = elt.ownerDocument.parentView;
    } ],
  tagName : ``p''
}
\end{lstlisting}
It is malicious because the ADsafe runtime could be tricked into
calling \lstinline|appendChild| with a direct DOM reference.
Fortunately, this is rejected by both JSLint and by a \tlint{}
checker, but for very different reasons.

JSLint rejects it because \lstinline|__nodes__| is a banned field.  In
contrast, the \tlint{}-checker assumes that 
the variable \lstinline|elt| is \tlint{}-typed.
This is sufficient to type-check the body of the function, whose type
(\type{Undef}) is subsumed by
\tlint{}.  Hence, the function has type $\tlint{} \rightarrow
\tlint{}$, which is subsumed by \tlint{}. 
Similarly, the \lstinline|__nodes__| field types to \tlint{} and the
object literal has type:
\begin{lstlisting}
{
  __nodes__ : $\tlint{}$,
  tagName : $\tlint{}$
}
\end{lstlisting}
However, the object literal, since it is a subexpression, must be
\tlint{}-typed too.  \tlint{} requires the object's
\lstinline|__nodes__| field to have type \tarray{\tdom}, which does
not match the \tlint{} type calculated above.  Hence, we have a type
error and the widget is rejected.

Relative to standard type machinery, JSLint is much more ad hoc.  For
instance, it rejects harmless programs whose variables happen to be
banned names, because it has no contextual information.  Thus, there
are safe widgets that the \tlint{}-checker accepts that JSLint
rejects.  Worse, the ad hoc nature of JSLint results in errors, such
as the missing static check in section~9.

\section{Unverifiable Code\label{sec:unverif-details}}

\begin{figure}
\begin{lstlisting}
var reject_name =  function (name) {
    return 
    ((typeof name !== 'number' || name < 0) &&
     (typeof name !== 'string' || 
      name.charAt(0) === '_' ||
      name.slice(-1) === '_' ||
      name.charAt(0) === '-'))
        || banned[name];
});

function F() {} // only used below

ADSAFE.create = 
  typeof Object.create === 'function'
  ? Object.create
  : function(o) {
      F.prototype = 
        typeof o === 'object' && o 
               ? o : Object.prototype;
      return new F();
    };
\end{lstlisting}
\caption{The Unverified Portion of ADsafe}
\label{fig:unverified}
\medskip\hrule
\end{figure}

Figure~\ref{fig:unverified} presents the entire unverified codebase
(re-formatted for column width), which we now discuss.

\paragraph{\texttt{reject\_name}} The \lstinline|reject_name| function
returns \lstinline|true| if its argument is a blacklisted field and
\lstinline|false| otherwise.  With a little work, we can give it a
type like \type{String} $\rightarrow$ \type{Bool}.  However,
\lstinline|reject_name| is used in predicate positions to guard
against invalid field accesses; therefore, it needs to return a more
precise type like
\begin{displaymath}
\type{UnsafeField} \rightarrow \type{True} \cap 
\type{SafeField} \rightarrow \type{Bool}
\end{displaymath}
where \type{UnsafeField} and \type{SafeField} are, respectively,
string-set types of blacklisted names and their complement.  The above
type lets us conclude that if the predicate returns false, the
argument was \emph{not} a blacklisted field, and can thus be
dereferenced safely.  The if-splitter
(section 7) uses this information.

Unfortunately, checking that the body of \lstinline|reject_name| has
this type is too sophisticated for our if-splitter.  To type it, we
would need a solver that incorporates the semantics of \code{charAt}
and other primitives.  Since we lack that, we use a \code{cheat} to
ascribe this type, and verify it by manual inspection.

\paragraph{\code{ADSAFE.create}}

In the (new) ECMAScript 5 standard, 
\lstinline|Object.create| takes an object \lstinline|o|
as a parameter and creates a new object whose prototype is
\lstinline|o|; if \lstinline|o| is not an object, the new object's
prototype is \lstinline|Object.prototype|.  ADsafe provides this same
functionality for current browsers through \lstinline|ADSAFE.create|.
This function is never used by ADsafe; it is only intended for
widgets.  Therefore, its type must be
\begin{displaymath}
[\twindow{} \cup \tlint{}] \tlint{} \cdots \rightarrow \tlint
\end{displaymath}
JSLint ensures that the actual argument is \tlint{}-typed
(section~5). However, the return type is problematic.
In our \tlint{} type (figure~4), the $proto$ field
admits \type{Object} but not \tlint{}, which is necessary to
type-check the code.  Permitting $\alpha$ (which represents \tlint{})
in the type of $proto$ results in a type system that we have not been
able to show will terminate.

\paragraph{\code{ADSAFE._intercept}}

ADsafe enables the hosting Web page to provide \emph{interceptors},
which are functions that get direct access to the DOM.  The above
count excludes interceptors because, by definition, these are unsafe.
Verifying interceptors requires analyzing the whole of the
\emph{page}, including its \html{} and how it modifies \as{}, which
are outside the scope of our work.

\end{document}